\documentclass[aps,prl,reprint,superscriptaddress]{revtex4-2}
 \usepackage{graphicx} 

\begin{document}


\title{On the scaling and anisotropy of two subranges in the inertial range of solar wind turbulence}


\author{Honghong Wu} 
\email{honghongwu@whu.edu.cn}
 \affiliation{School of Electronic Information, Wuhan University, Wuhan, People's Republic of China}
   
 
\author{Jiansen He}
\affiliation{School of Earth and Space Sciences, Peking University, Beijing, People's Republic of China} 

 \author{Liping Yang}
\affiliation{SIGMA Weather Group, State Key Laboratory for Space Weather, National Space Science Center, Chinese Academy of Sciences,Beijing, People's Republic of China} 

\author{Xin Wang}
\affiliation{School of Space and Environment, Beihang University, Beijing, People's Republic of China} 

\author{Shiyong Huang}
\affiliation{School of Electronic Information, Wuhan University, Wuhan, People's Republic of China} 

\author{Zhigang Yuan}
\affiliation{School of Electronic Information, Wuhan University, Wuhan, People's Republic of China} 

\date{\today}

\begin{abstract}
Intermittency and anisotropy are two important aspects of plasma turbulence, which the solar wind provides a natural laboratory to investigate. However, their forms and nature are still under debate, making it difficult to achieve a consensus in the theoretical interpretation. Here, we perform higher-order statistics for the observations in the fast solar wind at 1.48 au obtained by Ulysses and in the slow solar wind at 0.17 au obtained by Parker Solar Probe (PSP). We find that two subranges clearly exist in the inertial range and they present distinct features with regard to the intermittency and anisotropy. The subrange 1 with smaller scale has a multifractal scaling with the second index $\xi(2) \sim 2/3$ and the subrange 2 with larger scale is also multifractal but with $\xi(2) \sim 1/2$. The break between two subranges locates at the same spatial scale for both Ulysses and PSP observations. Subrange 1 is multifractal in the direction perpendicular to the local magnetic field with $\xi_{\perp}(2) \sim 2/3$ and seems to be monoscaling in the parallel direction with $\xi_{\parallel}(2) \sim 1$. Subrange 2 is multifractal in both parallel and perpendicular directions with $\xi_{\perp}(2) \sim 1/2$ and $\xi_{\parallel}(2) \sim 2/3$. Both subrange 1 and subrange 2 present power and wavevector anisotropies. The distinct features of two subranges suggest that a transition from weak to strong turbulence may occur and the spatial scale of the break may not evolve with the solar wind expansion. These new results update our knowledge of the inertial range and provide strong observational constraints on the understanding of intermittency and anisotropy in solar wind turbulence.
 
\end{abstract}


\maketitle

\section{Introduction}
Solar wind is a natural laboratory to investigate plasma turbulence, in which the intermittency and anisotropy are two critical aspects of the underlying nonlinear physics \cite{Bruno2013LRSP, Horbury2012SSRv}. Intermittency and anisotropy in solar wind turbulence play important roles in the energy transfer and cosmic-ray transport in the heliosphere. Great efforts are made to uncover their forms and nature by means of both theoretical \cite[e.g.][]{Goldreich1995ApJ, Boldyrev2006PhRvL, Zank2017ApJ} and observational studies \cite[e.g.][]{Burlaga1991GeoRL,Horbury2008PhRvL,Salem2009ApJ,Chen2012ApJ,Osman2014ApJL, Wang2014ApJ,He2019ApJ}. However, the precise forms are still under debate and their nature remains poorly understood. 
 
Theoretical models to describe the solar wind turbulence can be divided into two main classes. One is the 2D+slab and the other is the critical balance. Both offer plausible explanations for the observed anisotropies and propose the possible break in the inertial range. In the former framework, double power law may appear with a Kolmogorov-like \cite{Kolmogorov1941ANSD} index at smaller wavenumber regime dominated by quasi-2D fluctuations and an IK-like \cite{Iroshnikov1964SvA} index at larger wavenumber regime dominated by wave-like fluctuations \cite{Zank2020aApJ}. In the latter case, there may exist a transitional scale beyond which weak wave turbulence drives itself into the critical balance (strong) regime \cite{Goldreich1995ApJ}. Such a transition inside the inertial range is presented in a simulation of incompressible balanced decaying MHD turbulence and is argued to be a universal property of several anisotropic turbulent systems \cite{Meyrand2016PhRvL}.

However, it is not easy to distinguish the break inside inertial range from the power spectra. In fact, the inertial range is believed to be single power law for decades based on the widely applied spectral methods. Up to present, only two reports pointed out the possible presence of two subranges in the inertial range on the observed power spectra. One is for the fast solar wind at near-Earth region measured by WIND \cite{Wicks2011PhRvL}, but do not receive enough attention thereafter due to the limitation of large errors. The other is for the near-Sun solar wind from the same flow tube measured by PSP \cite{Telloni2022FrASS}, but only the indices of the omnidirectional spectrum are given. More convincing evidences are required to confirm the existence of two subranges in the solar wind.
    
The higher-order statistics, however, provides a promising way to unravel the scaling laws features hidden to the usual spectral methods \cite{Bruno2013LRSP}. In this Letter, we utilize the higher-order statistics to analyze the inertial range both in the fast solar wind at 1.48 au and in the slow solar wind at 0.17 au. We present the clear existence of two subranges with distinct scaling laws in the inertial range of solar wind turbulence for the first time and find that their physical spatial scales show little radial variation.

\section{Data}
The Ulysses spacecraft measures the magnetic field $\vec{B}$ by Vector Helium Magnetometer \cite{Balogh1992AA} with time resolution $ 1$ second and the plasma by Solar Wind Observations Over the Poles of the Sun \cite{Bame1992AA}. Time interval is taken within a fast solar wind at $r \sim 1.48 $ au and latitude $\phi \sim 47^\circ $ from 1995 May 1 to 1995 May 10, in which the flow speed $V_\mathrm{0}= 771 $ km/s, proton beta $\beta = 1.20$, and cross helicity $\sigma_c = -0.75$. We also use magnetic field data with time resolution $ 0.8738$ second obtained from the fluxgate magnetometer in the FIELDS instrument suite \citep{Bale2016SSRv} and plasma data obtained from the the Solar Probe Cup \cite{Case2020ApJS} of Solar Wind Electrons, Protons, and Alphas instrument suite \cite{Kasper2016SSRv} on board PSP. Time interval is taken during the first encounter phase from 2018 November 6 to 2018 November 7 when $r \sim 0.17 $ au, $V_\mathrm{0}= 332 $ km/s, $\beta = 0.54$, and $\sigma_c = 0.84$.

\section{MULTI-ORDER STRUCTURE FUNCTIONS}
The magnetic field increments are calculated as 
\begin{equation}
\delta \vec{B}(t,\tau)=\vec{B}(t)-\vec{B}(t+\tau),
\end{equation}    
where $\tau$ is the time lag. The q-th order magnetic-trace structure functions are defined as 
\begin{equation}
S^q (\tau)= <|\delta \vec{B}(t,\tau)|^q>,
\end{equation}    
where $<>$ denotes an ensamble time average. We calculate the local background magnetic field $\vec{B_\mathrm{0}} (t,\tau)$ as averages in the moving scale-dependent window $\tau$. $\theta (t,\tau)$ is obtained between the local magnetic field direction and the sampling direction (local velocity direction for PSP and radial direction for Ulysses). $\delta \vec{B}(t,\tau)$ are selected under the criteria of different sampling angle ($80^\circ<\theta(t,\tau)<100^\circ$ and $\theta(t,\tau)<10^\circ\ or\ \theta(t,\tau)>170^\circ$) to obtain $S^q (\tau_{\perp})$ and $S^q (\tau_{\parallel})$. For single spacecraft observations, the spatial lag $L=\tau \vec{V_\mathrm{0}}$ under Taylor hypothesis \citep{Taylor1938RS}. It is further normalized by the ion inertial length $d_i$ to reveal the underlying processes on a physically constant scale.
  
\begin{figure*}[ht!] 
\includegraphics[width=1.0\linewidth]{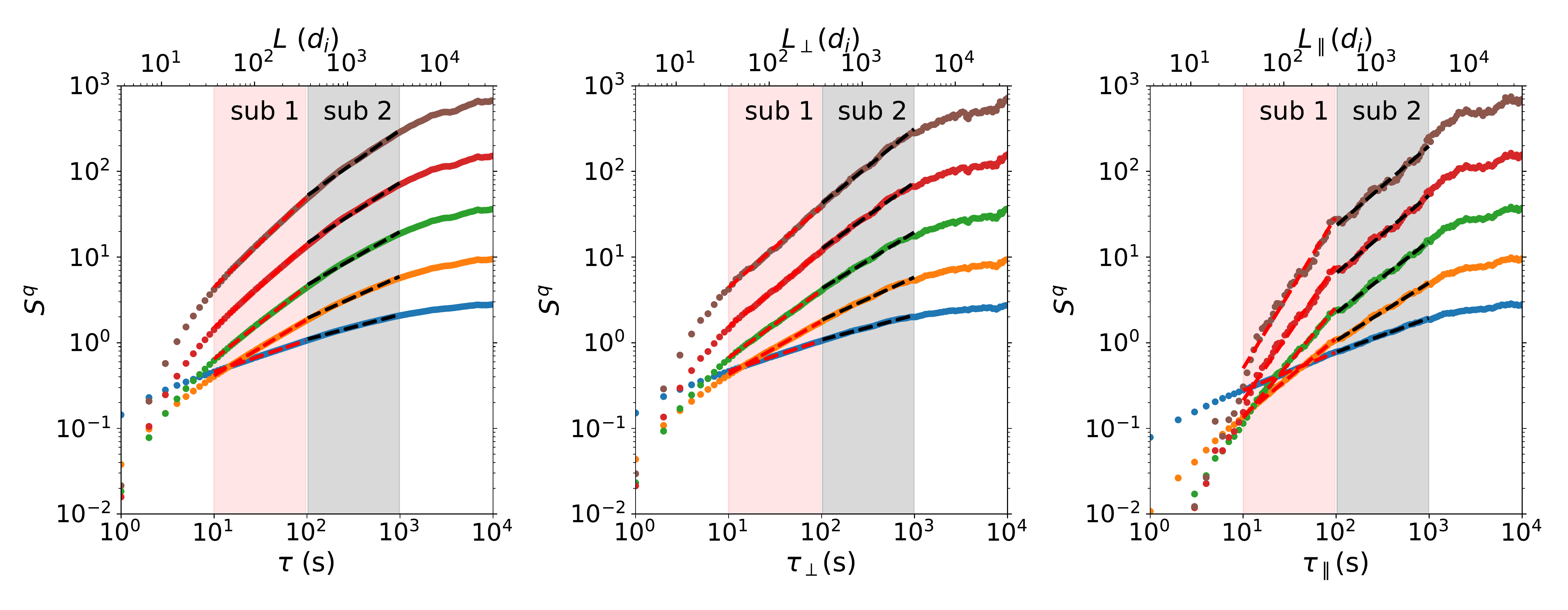} 
\caption{\label{fig:fig1} Magnetic-trace structure functions $S^q$ as functions of time lags $\tau$ (bottom axis) and spatial lags $L$ (upper axis, in unit of the ion inertial length $d_i$) in the fast solar wind at $1.48$ au observed by Ulysses. The blue, orange, green, red and purple represent the $S^q$ with order $q= 1, 2, 3, 4, 5$, respectively. The red and black dashed lines are power-law fits to $S_q$ in the corresponding subrange 1 and subrange 2, shown in red and black shadows respectively. Left: $S^q$ measured in the radial direction; middle and right: $S^q$ measured in the direction perpendicular and parallel to the local magnetic field.}
\end{figure*}

Fig. \ref{fig:fig1} presents $S^q (\tau)$, $S^q (\tau_{\perp})$, and $S^q (\tau_{\parallel})$ for the fast wind interval observed by Ulysses. It is obvious that $S^q (\tau_{\parallel})$ shows two distinct subranges $10<\tau<100$ s and $100<\tau<1000$ s within the normally considered inertial range determined by the power spectra (not shown). We refer to $10<\tau<100$ s as subrange 1 and $100<\tau<1000$ s as subrange 2 for this interval, corresponding to a spatial scale of $\sim 36-360\ d_i$ and $\sim 360-3600\ d_i$. The break locates around $ 100$ seconds and becomes clearer as the order goes higher. It is not easy to distinguish those two subranges from $S^q (\tau)$ and $S^q (\tau_{\perp})$, which is the very reason why the differences between them are easily overlooked in previous works.

$S^q (l)$ exhibits the scaling behaviour as  
\begin{equation}\label{equa:scaling}
S^q (l)= a_q l^{\xi(q)}.
\end{equation}  
Equation \ref{equa:scaling} contains full information of the specific range and $\xi(2)$ is related to the spectral index $\alpha$ by $\alpha = -\xi(2)-1$ \cite{Monin1975mit}. The dependence of $\xi$ on $q$ identifies the self-similarity or the intermittency of the turbulent system \cite{Biskamp2003}. A linear dependence is called monoscaling, implying a globally scale-invariant process. Nonlinear behaviour reflects multifractal, indicating that the energy distribution is not uniform space-filling but intermittent. Here we perform power law fits and obtain $\xi(q)$, $\xi_{\perp}(q)$, $\xi_{\parallel} (q)$ from $S^q (\tau)$, $S^q (\tau_{\perp})$, and $S^q (\tau_{\parallel})$ for subrange 1 and subrange 2, respectively. We find distinct features between two subranges.
  
\begin{figure*}[ht!] 
\includegraphics[width=1.0\linewidth]{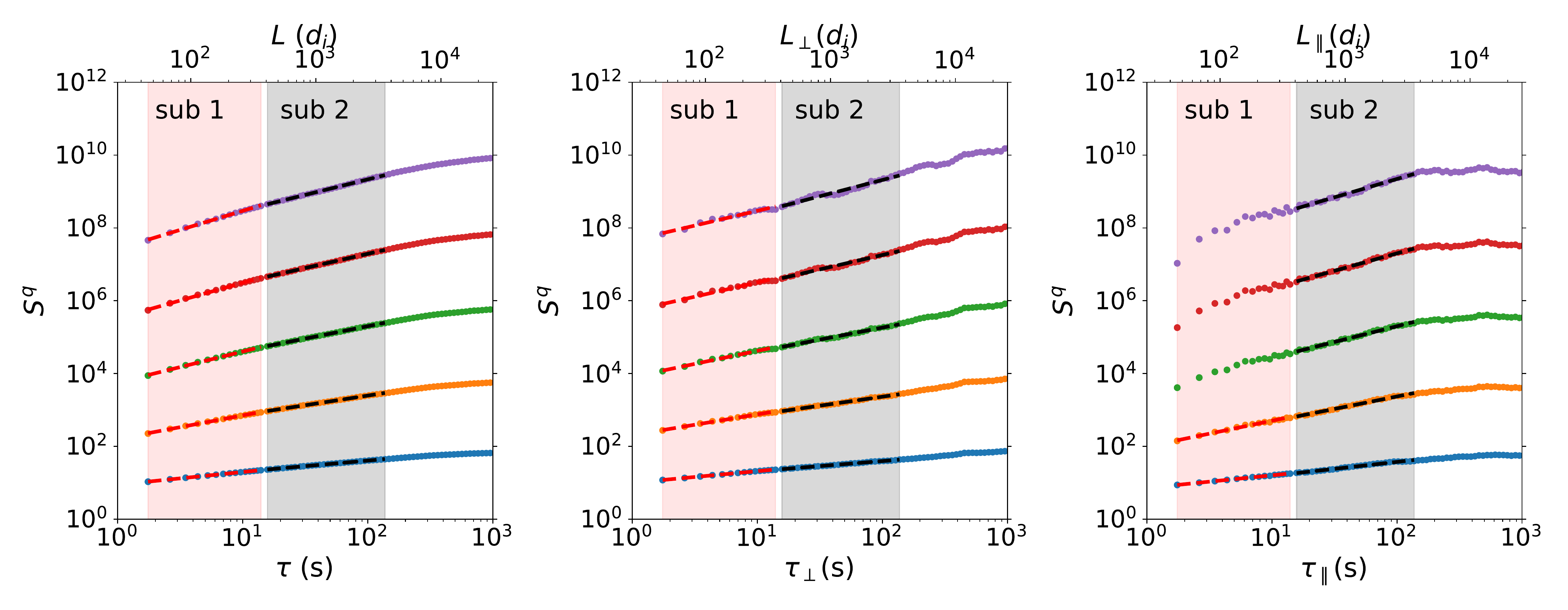} 
\caption{\label{fig:fig2} Magnetic-trace structure functions $S^q$ in the slow solar wind at 0.17 au observed by PSP with the same formats as Fig. \ref{fig:fig1}.
}
\end{figure*}

Fig. \ref{fig:fig2} presents $S^q (\tau)$, $S^q (\tau_{\perp})$, and $S^q (\tau_{\parallel})$ in the same manner as Fig. \ref{fig:fig1} for the PSP observation. We divide the ranges into subrange 1 and subrange 2 according to the spatial scale $L$. The red shadows in Fig. \ref{fig:fig1} and Fig. \ref{fig:fig2} share the same spatial scales, so do the black shadows. The time scales of subrange 1 and subrange 2 for this interval at 0.17 au are around $1.5<\tau<15$ s and $15<\tau<150$ s, approximately an order lower than those for the interval at 1.48 au observed by Ulysses. We again perform power-law fits for those structure functions in the respective subrange 1 and subrange 2. 

\section{scaling and Anisotropy}

To guarantee the meaningful estimation of higher-order moments from the measurements, we follow the procedure in \cite{dewit2004PhRvE} to determine the trustworthy maximum order $q_{max}$, which is described in detail in the appendix. 
Fig. \ref{fig:fig3} presents $\xi(q)$, $\xi_{\perp}(q)$, $\xi_{\parallel}(q)$, in which the indices with higher than $q_{max}$ are marked in hollow style. Standard errors of these indices are small and no larger than the size of the markers. It clearly demonstrates the differences in terms of the scaling behaviour and its anisotropy between subrange 1 and subrange 2.  Fig. \ref{fig:fig3} (a) shows that $\xi$ from Ulysses and PSP observations are surprisingly overlapped in subrange 1 from the 1st to the 5th order, illustrating the multifractal scaling and $\xi(2) \sim 2/3$. They are similar to each other in subrange 2, demonstrating again the multifractal but $\xi(2)$ now close to $1/2$. The transition from $2/3$ to $1/2$ is consistent with the transition of the power spectral indices \cite{Telloni2022FrASS}. Both the difference between two subranges and the similarity between two intervals prove the existence of two distinct subranges in the inertial range of solar wind turbulence.

\begin{figure*}[ht!] 
\includegraphics[width=1.0\linewidth]{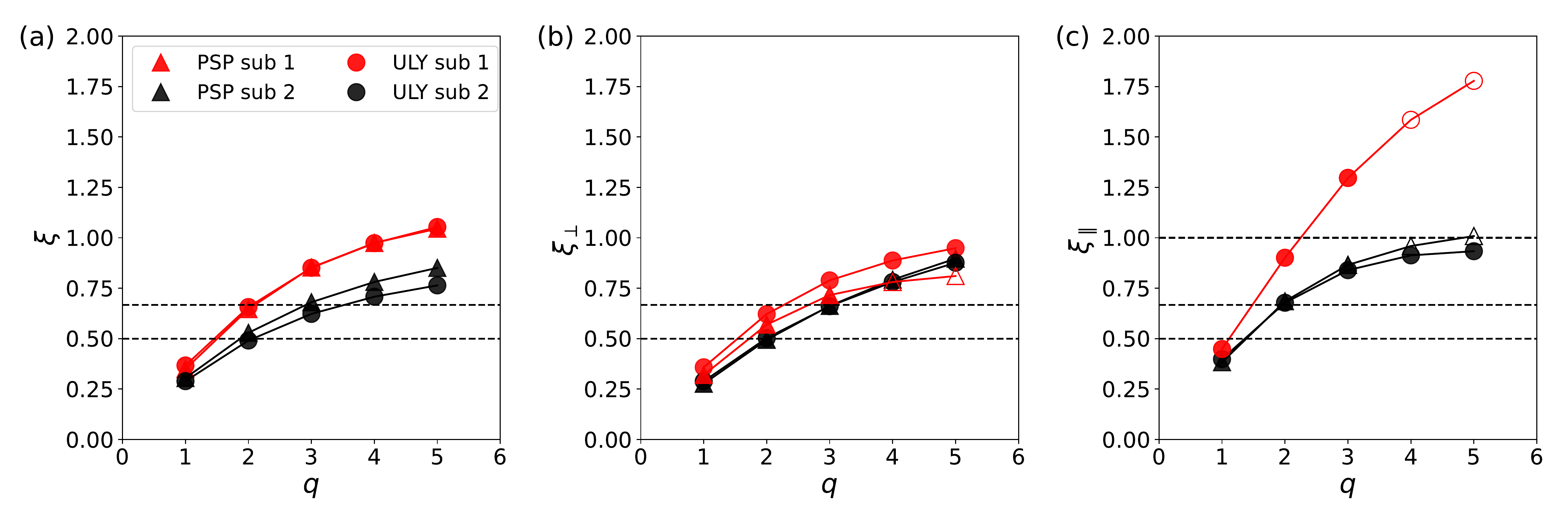} 
\caption{\label{fig:fig3} $\xi$ (left), $\xi_{\perp}$ (middle), $\xi_{\parallel}$ (right) as a function of the order $q$. Triangles represent PSP observations and dots represent Ulysses observations. The red and black represent the results in subrange 1 and subrange 2, as denoted by the red and black shadows in Fig. \ref{fig:fig1} and \ref{fig:fig2}. The dashed horizontal lines denotes $\xi = 1/2,2/3,1$. Those indices with higher than $q_{max}$ are marked in hollow style.
}
\end{figure*}

The anisotropy of the scaling is also investigated. Fig. \ref{fig:fig3} (b) shows the scaling observed in the direction perpendicular to the local magnetic field. The Ulysses and PSP observations from the 1st order to the 5th order are nearly overlapped in subrange 2 and they are adjacent to each other in subrange 1. Note that the order higher than 3 are unreliable for PSP observations of the subrange 1, so that the increasing gap between the PSP interval and Ulysses interval may not be physically true. The overall difference between two subranges is smaller than that in Fig. \ref{fig:fig3} (a). However, the same features can be captured that both subranges are multifractal and $\xi_{\perp}(2)$ is close to $1/2$ in subrange 2.

Fig. \ref{fig:fig3} (c) presents the scaling detected in the direction parallel to the local magnetic field. It is clear that in subrange 2, PSP interval and Ulysses interval behave similarly with a multifractal scaling and $\xi(2) \sim 2/3$. Since the trustworthy highest order is only 1st for PSP in subrange 1, we do not show the result. However, from the Ulysses observation, a linear monoscaling can be identified with $\xi_{\parallel}(2)$ closer to $1$.  

\begin{figure*}[ht!] 
\includegraphics[width=1.0\linewidth]{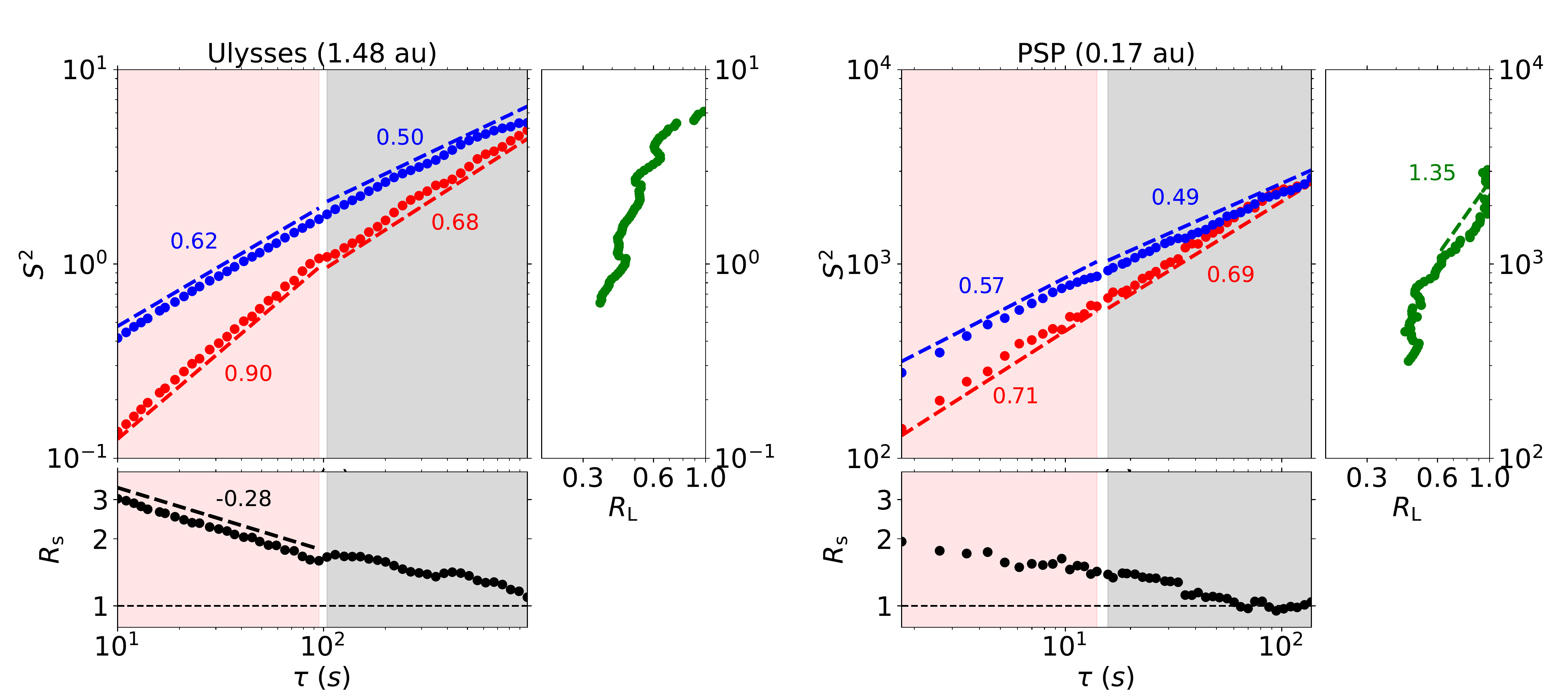} 
\caption{\label{fig:fig4} Anisotropies of scaling index, power, and wavevector revealed by $S^2$ observed by Ulysses (left) and PSP (right). Red and black shadows denote subrange 1 and subrange 2. Blue and red dots represent $S^2(\tau_{\perp})$ and $S^2(\tau_{\parallel})$. Black and green dots represent the power and wavevector anisotropies, respectively. The numbers with the colors show the power law fitted indices for the corresponding dots and the dashed lines with these indices are plotted to guide the eye. 
}
\end{figure*}

Fig. \ref{fig:fig4} emphasizes the 2rd order structure functions and three associated anisotropies of spectral index, power, and wavevector, which can be inferred by power spectra as well. In the left panels, the 2rd order structure functions of the Ulysses interval display two subranges with a distinct break. Subrange 1 shows $\xi_{\parallel}(2) =0.90$ and $\xi_{\perp}(2)=0.62$, while Subrange 2 shows $\xi_{\parallel}(2)=0.68$ and $\xi_{\perp}(2)=0.50$. The spectral index anisotropy in subrange 1 are similar with previous works found using spectral method \cite{Wicks2011PhRvL}. However, it is the first time to distinguish the spectral indices in subrange 2 with $\xi_{\parallel}(2)\sim 2/3$ and $\xi_{\perp}(2) \sim 1/2$.

In the right panels of Fig. \ref{fig:fig4}, the 2rd order structure functions of the PSP interval are analyzed. Unfortunately, the parallel structure functions higher than 1 in subrange 1 is not trustworthy. We focus on subrange 2 here. The spectral index anisotropy for the near-Sun solar wind in the inertial range around $10-100$ seconds shows that $\xi_{\parallel}(2) =0.69$ and $\xi_{\perp}(2)=0.49$, sharing exactly the same feature as that of subrange 2 for the Ulysses interval.
 
The black dots in both lower panels of Fig. \ref{fig:fig4} present the power anisotropy by $R_\mathrm{S}$ versus $\tau$. $R_\mathrm{S}(\tau) = S^2(\tau_{\perp}=\tau)/S^2(\tau_{\parallel}=\tau)$. $R_\mathrm{S}$ decreases as $\tau$ increases and becomes near 1 as the $\tau$ approaches the boundary of the inertial range.  $R_\mathrm{S}(\tau) \propto \tau^{-0.28} $ in subrange 1. The green dots in both right panels draw the wavevector anisotropy by $R_\mathrm{L}$ versus $S^2$. $R_\mathrm{L}(S^2) = L(S^2(L_{\perp})=S^2)/ L(S^2(L_{\parallel})=S^2)$. $R_\mathrm{L}$ increases as $S^2$ increases and becomes near 1 as the power level arrives at the boundary of the inertial range. $R_\mathrm{L}(S^2) \propto (S^2)^{1.35} $ in subrange 2.

\section{Conclusions and discussion}

We perform the multi-order structure function analyses for the magnetic field and find two distinct subranges in the inertial range of plasma turbulence, not only in the near-Sun solar wind but also in the fast solar wind at 1.48 au. We find that both subranges exhibit multifractal scalings, but subrange 1 has the 2rd scaling index close to the predicted $2/3$ in Komolgorov-like turbulence, while subrange 2 close to the predicted $1/2$ in Iroshnikov-Kraichnan phenomenology. 

We present the anisotropy of the multi-order scaling behaviour for subrange 2 to subrange 1 respectively for the first time. Subrange 1 is multifractal in perpendicular direction but monofractal in parallel direction. However, we find that subrange 2 is multifractal in both parallel and perpendicular direction. It is the first time to show the multi-order scaling anisotropy for the near-Sun solar wind using PSP observations. We also show the scaling index, power and wavevector anisotropy for both subranges. It is the first time to distinguish $2/3$ and $1/2$ for the scaling index of subrange 2. Both power anisotropy and wavevector anisotropy indicate that it is isotropic at the outer scale and the anisotropy develops as energy cascades towards smaller scale in the inertial range.

Previous works reported that the spectral index evolves from $-3/2$ to $-5/3$ in the inertial range from the near-Sun region to 1 au observed by PSP \cite{Chen2020ApJS,Alberti2020ApJ} and the fluctuations exhibit multifractal scaling without radial evolution \cite{Zhao2020ApJ, Chhiber2021ApJ, Sioulas2022ApJ}. The observations in the near-Earth solar wind show that the magnetic field fluctuations turns from monoscaling to multifractal scaling \cite{Osman2014ApJL} and are with the spectral index from near $-2$ to $-5/3$ when measuring from parallelly to perpendicularly to the local magnetic field \cite{Horbury2008PhRvL, Podesta2009ApJ, Luo2010ApJL}. The observations in the near-Sun solar wind, however, show the spectral indices close to $-5/3$ in the parallel direction and near $-3/2$ in the perpendicular direction \cite{Huang2022ApJL}. Here we show that the differences between the near-Sun and near-Earth solar wind probably result from the existence of two subranges in the inertial range and the evolution of ion inertial length $d_i$, which make the frequency range to be analyzed move from a physically larger range to a physically smaller range, that is, from subrange 2 ($\sim 30\ d_i-300\ d_i$) to subrange 1 ($\sim 400\ d_i-4000\ d_i$). Subrange 1 and 2 appear at both 1.48 and 0.17 au without a clear change of the transitional spatial scale, suggesting that the evolution may be intrinsic inside the inertial range, rather than a passive behaviour due to the solar wind expansion. Future statistical studies are required to answer the question whether the break is a universal property or not.
  
In summary, we present clearly the existence of two distinct subranges and their respective scaling and anisotropy in the inertial of solar wind turbulence using higher-order statistics for the first time. Our new findings give a delicate form of the intermittency and anisotropy in solar wind turbulence and provide new insights into the nature of turbulent fluctuations. The two subranges in the inertial range certainly deserve further investigation. A spectral subrange called the intermittency domain in the inertial range is found \cite{Wang2014ApJ} and the relationship between this domain and the two subranges awaits future study.

\appendix*

\begin{acknowledgments}
We acknowledge the PSP and Ulysses mission for use of the data, which are publicly available at the NASA Space Physics Data Facility (https://spdf.gsfc.nasa.gov/). This work is supported by the National Natural Science Foundation of China under contract Nos. 42104152, 41974198,42174194,41874200,41874199 and by CNSA under contract Nos. D020301, D020302, and D050601, as well as National Key R\&D Program of China No. 2021YFA0718600. X. Wang is also supported by the Fundamental Research Funds for the Central Universities of China (KG16152401, KG16159701).  
\end{acknowledgments}

\section{Determine the maximum reliable order}
The magnetic field increments $\delta \vec{B}(t,\tau)$ are selected under the criteria of different sampling angle ($0^\circ<\theta(t,\tau)<180^\circ$, $80^\circ<\theta(t,\tau)<100^\circ$ and $\theta(t,\tau)<10^\circ\ or\ \theta(t,\tau)>170^\circ$) to obtain $S^q (\tau)$, $S^q (\tau{\perp})$, and $S^q (\tau_{\parallel})$. The absolute values of the selected magnetic field increments $\delta \vec{B}(t,\tau)$ are sorted in decreasing order to be $\delta \vec{B}(k,\tau)$ with the indices $k$ representing the position. The peak of $\delta \vec{B}(k,\tau)$ with respect to $k$ is well described by a power law $\delta \vec{B}(k,\tau) =\alpha k^{-\gamma(\tau)}$. We perform the power-law fits for different sets of $\delta \vec{B}(k,\tau)$ and obtain $\gamma(\tau)$.

\begin{figure*}[ht!] 
\includegraphics[width=1.0\linewidth]{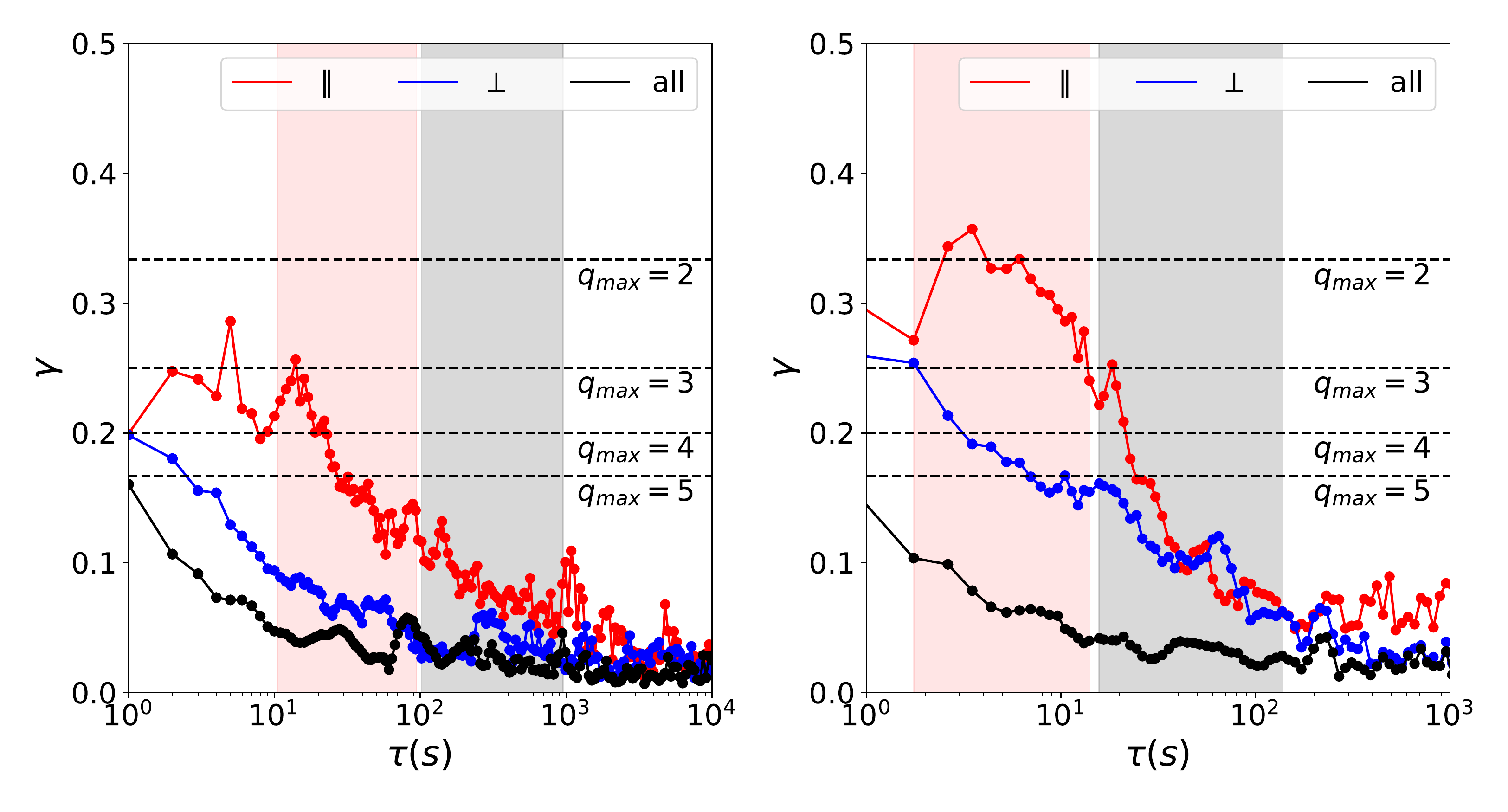} 
\caption{\label{fig:fig5} scaling exponent $\gamma$ obtained by the least-squares fit over the sorted magnetic field increments versus their index for the solar wind at 1.48 au observed Ulysses (left) and at 0.17 au observed by PSP (right).  The black, blue and red represent the measuring direction along with the solar wind velocity, perpendicular to the local magnetic field, and parallel to the local magnetic field, respectively. The red and black shadows denote subrange 1 and subrange 2. The horizotal dashed lines denote the position of $\gamma$ with the corresponding $q_{max}$ shown. 
}
\end{figure*}

The maximum moment $q_{max}$ that can be meaningfully estimated is determined by $\gamma$ \cite{dewit2004PhRvE}:  
\begin{equation}
q_{max} =[\frac{1}{\gamma}]-1,
\end{equation}
here $[]$ denotes the integer part. Fig. \ref{fig:fig5} presents $\gamma(\tau)$ for $S^q (\tau)$, $S^q (\tau_{\perp})$, and $S^q (\tau_{\parallel})$ respectively. $S^q (\tau)$ with $\gamma(\tau)$ lower than the horizotal dashed lines marked by $q_{max}$ can provide reliable results with $q<q_{max}$. Therefore, we can tell that $S^q (\tau)$ is reliable up to 5th order for both subrange 1 and 2. $S^q (\tau_{\perp})$ is reliable up to 5th order for subrange 2. $S^q (\tau_{\perp})$ is reliable up to 5th order for subrange 1 observed by Ulysses and only with $q<3$ for subrange 1 observed by PSP. $S^q (\tau_{\parallel})$ is reliable up to 5th order for subrange 2 and with $q<3$ for subrange 1 observed by Ulysses. $S^q (\tau_{\parallel})$ is reliable with $q<3$ for subrange 2 and only with $q<1$ for subrange 1 observed by PSP.

\end{document}